\documentclass[twocolumn,showpacs,pre,amsmath,amssymb]{revtex4}

\usepackage{graphicx}
\usepackage{dcolumn}
\usepackage{bm}

\begin{document}

\title{
High-resolution x-ray study of the nematic - smectic-A and smectic-A - smectic-C
transitions in $\overline{8}S5$-aerosil gels.
} 

\author{P.S.~Clegg}
\affiliation{Department of Physics, University of Toronto, Toronto, Ontario M5S 1A7, Canada}

\author{R.J.~Birgeneau}
\affiliation{Department of Physics, University of Toronto, Toronto, Ontario M5S 1A7, Canada}

\author{S.~Park}
\altaffiliation{Present Address: Center for Neutron Research, National Institute of
Standards and Technology, Gaithersburg, MD 20899}
\affiliation{Center for Material Science and Engineering, Massachusetts Institute of Technology, Cambridge, MA 02139}

\author{C.W.~Garland}
\affiliation{Center for Material Science and Engineering, Massachusetts Institute of Technology, Cambridge, MA 02139}

\author{G.S.~Iannacchione}
\affiliation{Department of Physics, Worcester Polytechnic Institute, Worcester, MA 01609}

\author{R.L.~Leheny}
\affiliation{Department of Physics and Astronomy, Johns Hopkins
University, Baltimore, MD 21218 - 2686}

\author{M.E.~Neubert}
\affiliation{Liquid Crystal Institute, Kent State University, Kent,
Ohio, 44242}

\date{\today}

\begin{abstract}
The effects of dispersed aerosil nanoparticles on two of the phase transitions of 
the thermotropic liquid crystal material 
4-$n$-pentylphenylthiol-4'-$n$-octyloxybenzoate ($\overline{8}S5$) 
have been studied using high-resolution x-ray diffraction techniques. The aerosils 
hydrogen bond together to form a gel which imposes a weak quenched 
disorder on the 
liquid crystal. The smectic-A fluctuations are well characterized by a two-component 
line shape representing thermal and random-field contributions. An elaboration on 
this line shape
is required to describe the fluctuations in the smectic-C phase; specifically the 
effect of the tilt on the
wave-vector dependence of the thermal fluctuations must be explicitly taken into
account. Both the magnitude
and the temperature dependence of the smectic-C tilt order parameter are observed to
be unaffected by the disorder. This may be a consequence of the large bare smectic 
correlation length in the direction of modulation for this transition.
These results show that the understanding developed
for the nematic to smectic-A transition for octylcyanobiphenyl (8CB) and
octyloxycyanobiphenyl (8OCB) liquid crystals with quenched disorder can be extended 
to quite different materials and transitions.
\end{abstract}

\pacs{64.70.Md,61.30.Eb,61.10.-i}

\maketitle

\section{Introduction}
\label{sec:intro}

The phase transition behavior of many thermotropic liquid crystals is quite well
characterized~\cite{deGennes_Prost} making liquid crystals ideal materials for studying the
effects of weak perturbations, including especially quenched random fields and 
interactions. In a recent series of papers~\cite{Park,Leheny,Germano2,Clegg}, it has 
been shown that the 
nematic to smectic-A (N - SmA) transition in liquid crystal-aerosil dispersions is
well described as a 
continuous symmetry (XY) transition with quenched random fields. This facilitates
studying this important class of transitions. Early calorimetric studies have
shown a systematic variation in the critical fluctuations as the aerosil
concentration increases. These experiments and accompanying theory have
focused on the isotropic to nematic (I - N)~\cite{Germano1} and the N - SmA
transitions~\cite{Park,Leheny,Germano2,Clegg,Germano1}. Here we present results on a
substantially different material at the N - SmA transition and we present the first
x-ray data for the
smectic-A to smectic-C (SmA - SmC) transition for liquid crystal-aerosil
dispersions. The SmA - SmC transition for heptyloxybenzylidene(7O.4)-aerosil 
dispersions has previously been 
investigated using ac-calorimetry~\cite{H_and_G}.
\begin{table*}
\caption{\label{tab:Tna} Characteristics of the aerosil gel structure and the 
$\overline{8}S5$-aerosil transitions. $\rho_S$ is the mass of aerosil divided
by the volume of liquid crystal; $\Phi$ is the volume fraction
of pores; $\ell_0 = 2/a\rho_S$ is the mean aerosil pore size, where $a$ is the specific
silica area~\cite{Germano1}; $\xi_{\|}^{NA}$, $\xi_{\bot}^{NA}$ indicate the approximate SmA
parallel and perpendicular correlation lengths close to the N - SmA 
pseudotransition~\cite{res_note} and 
$T_{AC}^*$ is the SmA - SmC transition temperature.
}
\begin{ruledtabular}
\begin{tabular}{c@{\extracolsep{0.5cm}}ccccc}
$\rho_S$ (g~cm$^{-3}$) & $\Phi$ & $\ell_0$ (\AA) & $\xi_{\|}^{NA}$ (\AA) & 
$\xi_{\bot}^{NA}$ (\AA) & $T_{AC}^*$ (K) \\
\hline
0.0 & 1 & $\infty$ & $\infty$ & $\infty$ & $329.3 \pm 0.1$ \\
0.025 & 0.99 & 2700 & $\sim$40000 & $\sim$1000 & $327.8 \pm 0.2$\\
 0.039 & 0.98 & 1700 & $\sim$16000 &$\sim$650 & $323.7 \pm 0.4$ \\
 0.073 & 0.97 & 910 & $\sim$15000 & $\sim$600 & $323.8 \pm 0.2$ \\
 0.096 & 0.96 & 690 & 8000 & 340 & $326.1 \pm 0.3$ \\
 0.220 & 0.91 & 300 & 2400 & 130 & $325.9 \pm 0.2$ \\
 0.350 & 0.86 & 190 & 4000 & 200 & $327.0 \pm 0.2$ \\
\end{tabular}
\end{ruledtabular}
\end{table*}

\begin{figure}[b]
\includegraphics[scale=0.3]{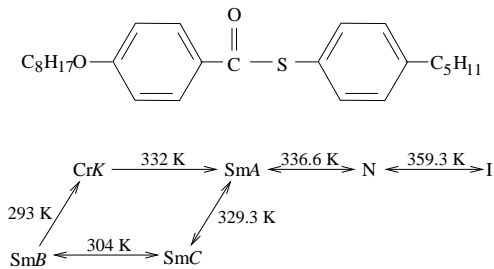}
\caption{\label{figure1} The chemical structure and phase sequence for bulk
$\overline{8}S5$. The transition temperatures are those given in 
Refs.~\cite{Schantz} and \cite{Neubert81}.}
\end{figure}

4-$n$-pentylphenylthiol-4'-$n$-octyloxybenzoate ($\overline{8}S5$) is a non-polar
liquid crystal with a succession of ordered phases (Fig.~\ref{figure1}). At 
$T_{NA}^0 = 336.6$ K a transition from
the orientationally ordered N phase to the layered SmA phase occurs. For this particular
liquid crystal the pretransitional correlation volume is highly
anisotropic~\cite{Safinya81}. As we shall demonstrate in the present work,
this gives rise to a strong
asymmetry in the powder-averaged line shape observed with x-ray diffraction.
This effect is enhanced by the
relatively small fourth-order correction to the correlation function in the transverse 
direction. Thus, the spherically averaged smectic 
line shape is very different from that
observed for materials such as octylcyanobiphenyl (8CB) and octyloxycyanobiphenyl 
(8OCB). Consequently, this $\overline{8}S5$ study provides a test of the generality of 
the line-shape analysis used in Refs. \cite{Leheny} and \cite{Clegg} versus that used in 
Ref. \cite{Aerogel_Science}. As the
temperature is lowered further, $\overline{8}S5$ exhibits a SmA - SmC 
transition ($T_{AC}^0 = 329.3$ K) where the molecules develop a tilt with respect to the
layer normal. The tilt angle is the order parameter for this phase transition. The
transitional behavior has mean-field rather than critical characteristics albeit
with an anomalously large sixth order term in the 
free energy~\cite{Birgeneau83},
which is typical for the Landau tricritical to mean-field crossovers observed for most
SmA - SmC transitions~\cite{M_and_G}. 

A liquid crystal is perturbed with weak-to-intermediate strength quenched disorder when aerosil
particles are introduced. These particles are $\sim7\,$nm diameter silica spheres with
hydrophilic surfaces. When aerosil is dispersed in a liquid crystal, hydrogen
bonds form between the aerosil particles creating a low density gel. On very long
length scales the gel structure exhibits fractal correlations~\cite{Germano1}; however,
the gel has no particular correlations commensurate with the wave vector of the SmA
structure and hence is effectively random for our purposes. Here we present high-resolution
x-ray diffraction
measurements for $\overline{8}S5$ samples containing a range of concentrations of
aerosil particles. Both the SmA and SmC type liquid crystal correlations in these samples have been
characterized as a function of temperature.  

Previous x-ray studies of the N - SmA transition in liquid crystal-aerosil dispersions
demonstrated that the transition responded to the aerosil as though its principal
effect was to apply a random
field that pinned the phase of the density wave~\cite{Park,Leheny,Clegg}. The static
random-field fluctuations are more effective than thermal fluctuations at destroying
long-range smectic order, and this was evident in the development of the structure
factor as a
function of temperature and aerosil concentration. Here we wish to investigate whether the
non-polar $\overline{8}S5$ is less strongly perturbed than the polar cyano
materials 8CB and 8OCB were. In
addition, the tilt order parameter at the mean-field SmA - SmC transition couples
to the aerosil in a different manner than the SmA order parameter.

Section~\ref{sec:exp} describes the experimental techniques used to make and study the
samples. In Section~\ref{sec:R_and_A} the analysis protocol is outlined and the data and
model parameters are presented. Finally in Section~\ref{sec:conc} we draw conclusions from
our measurements.

\section{Experimental techniques}
\label{sec:exp}

The $\overline{8}S5$ was synthesized at Kent State University and the transition temperatures
were observed to be close to those previously recorded for the bulk 
material~\cite{Schantz,Neubert81}. The aerosil
particles (hydrophilic type-300) were provided by Degussa and were dried prior to sample 
preparation. The
aerosil concentration was characterized by $\rho_S$, the mass of aerosil divided by
the volume 
of liquid crystal. Samples with $\rho_S = 0.0, 0.025, 0.039, 0.073, 0.096, 0.220,
$ and $ 0.350$~g~cm$^{-3}$ were prepared (we drop the units hereafter). The corresponding 
pore volume fractions and mean
pore sizes are given in Table~\ref{tab:Tna}. As with previous studies of 8OCB-aerosil
dispersions~\cite{Clegg}, it was necessary to
avoid crystallization of the $\overline{8}S5$ samples prior to measurement since
this results 
in phase separation.  The aerosil particles were dispersed in a solution of
$\overline{8}S5$ in spectroscopic grade acetone; the system was then
sonicated and kept well above the crystallization temperature until the
measurements were performed. The x-ray studies were undertaken once the acetone
solvent had been
evaporated. The sample environment and experimental geometry have been described in
detail previously~\cite{Clegg}. The studies were made using 8$\,$keV x-rays at the
X20A beam line at the National Synchrotron Light Source. Of great interest here was
the asymmetry of the smectic scattering peak and its evolution with temperature
and disorder strength. To 
insure that the observed effects were due to the sample and not the instrumental
resolution, measurements were made at both positive and negative scattering angles.
As with previous studies, the number of scans was kept limited to avoid undue radiation
damage to the sample. The results obtained in the N, SmA, and SmC regimes are reported
below.
\begin{figure}
\centerline{\includegraphics[scale=0.27]{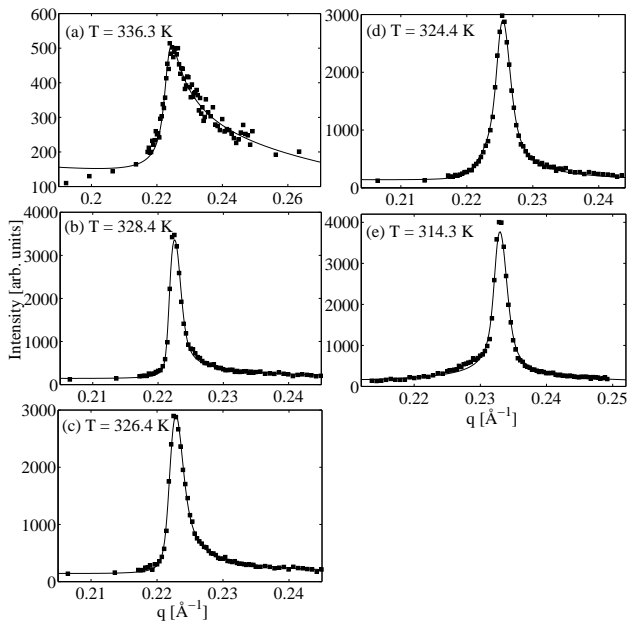}}
\caption{\label{figure2} (a-e) The x-ray intensity versus wave-vector transfer with
decreasing temperature for a $\overline{8}S5$-aerosil dispersion with $\rho_S =
0.096$~g~cm$^{-3}$. Panels (a-c) show the asymmetric 
line shape typical of SmA fluctuations in $\overline{8}S5$. The pseudotransition
into the short-range SmC state occurs at a temperature just below that for panel (c). The
solid lines are 
the results of fits described in the text.}
\end{figure}

\section{Results and analysis}
\label{sec:R_and_A}

Figure~\ref{figure2} shows a succession of scans of $\overline{8}S5$ with 
$\rho_S = 0.096$ as the temperature is decreased.
At high temperatures in the nematic phase (Fig.~\ref{figure2}(a)) there 
is a broad peak
corresponding to short range pretransitional SmA fluctuations. The observed powder average
has a characteristic asymmetry. On reducing the temperature the correlated regions grow in size
as indicated by the narrower line profile (Fig.~\ref{figure2}(b)). Unlike bulk 
$\overline{8}S5$,
the liquid crystal enters a SmA state with a finite correlation volume. Further cooling takes the
sample toward the SmC state. Prior to the onset of the molecular tilt, the peak begins to broaden
(Fig.~\ref{figure2}(c)). As the sample enters the SmC regime the peak moves to higher wave vectors
and begins to narrow (Fig.~\ref{figure2}(d)). At the lowest temperatures studied, the
angle of molecular tilt exceeds $20^{\circ}$ while the line-shape asymmetry is reversed
compared to that in the SmA phase (Fig.~\ref{figure2}(e)). The solid lines in Fig.~\ref{figure2} are
the results of fits to a model to be described below. 
\begin{figure}
\includegraphics[scale=0.4]{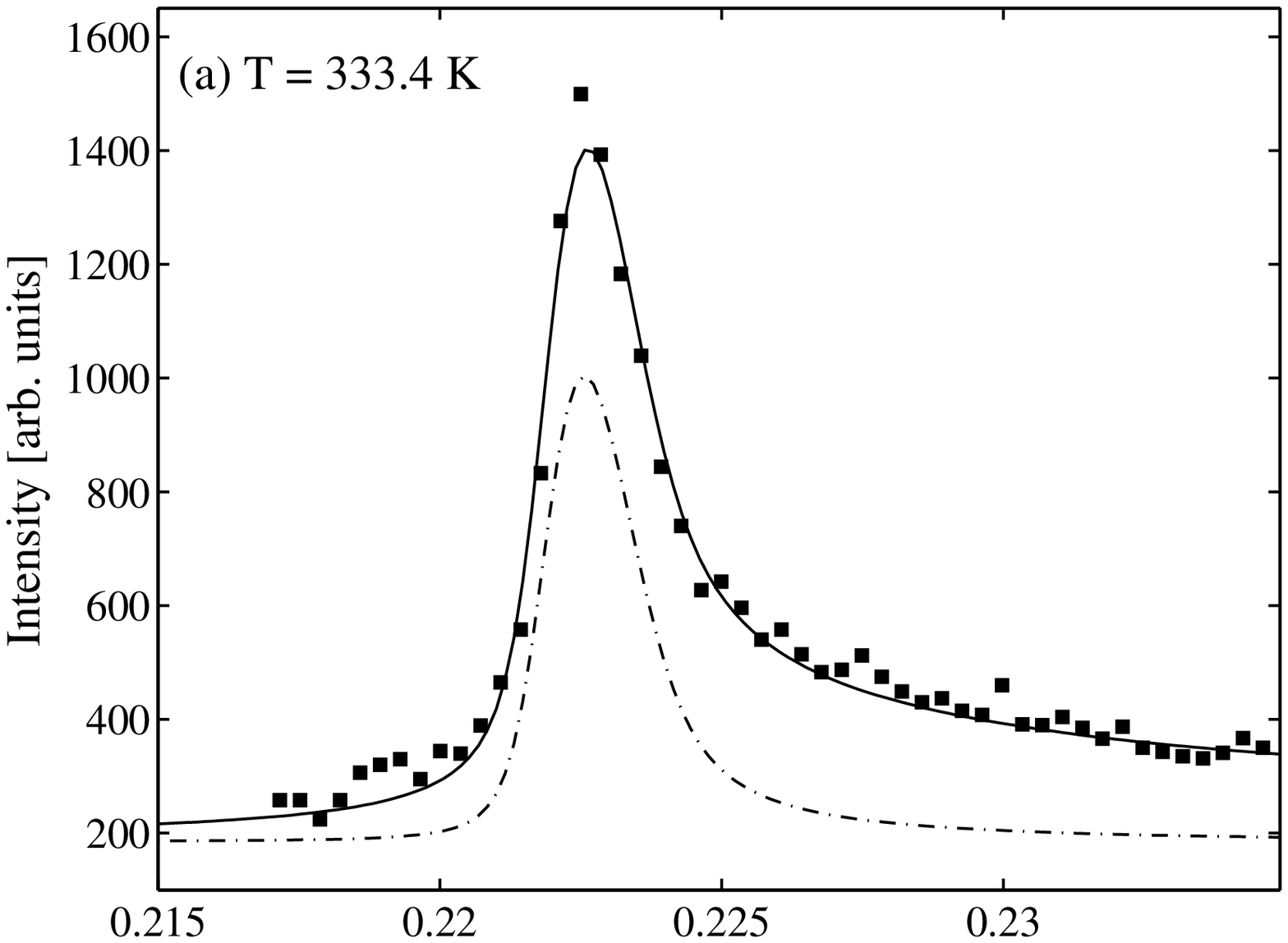}
\includegraphics[scale=0.4]{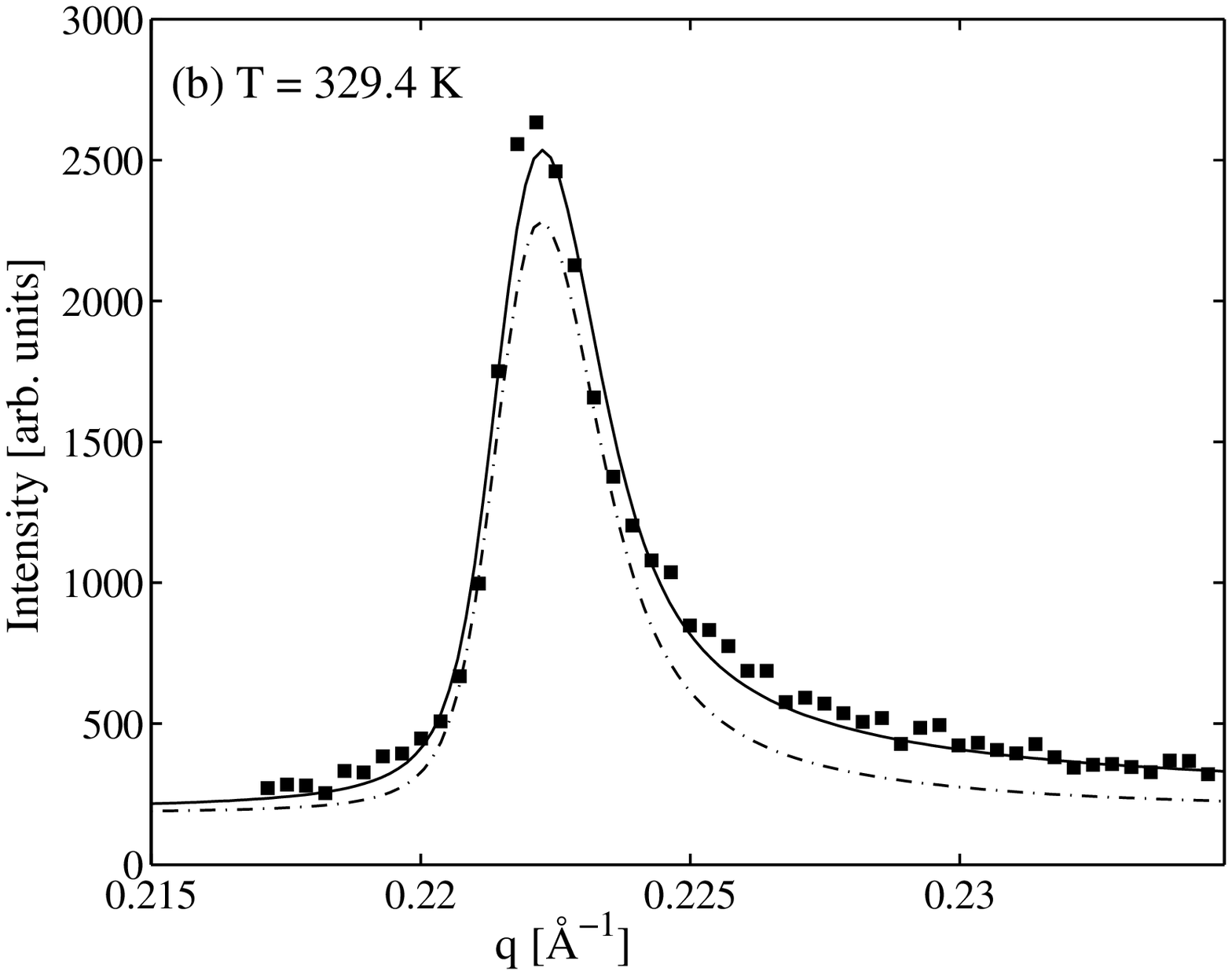}
\caption{\label{figure3} The x-ray intensity in reciprocal space for a
$\overline{8}S5$-aerosil dispersion with $\rho_S =
0.22$~g~cm$^{-3}$ at two temperatures in the SmA regime. The full line is the result
of a fit involving the powder average of Eq.~(\ref{N_to_A}) to the data. The broken
line represents only the second term corresponding to the random-field fluctuations. The
random-field contribution increases with decreasing temperature
yielding a characteristic change in the observed line shape.} 
\end{figure}

In keeping with previous studies of the
N-SmA transition in liquid crystal - aerosil dispersions~\cite{Leheny,Clegg}, the
line shape has been taken to be 
composed of two terms: the first is based on the line shape for the SmA thermal
fluctuations in the pure liquid crystal and the second is the same line shape
squared, which corresponds to the contribution of static random-field fluctuations. The
latter are caused by the 
randomly positioned aerosil surfaces pinning the phase of the density wave. The results of
fits in Fig.~\ref{figure2}(a-c) and in Fig.~\ref{figure3} show that this line shape describes
the data in the SmA region very well. The full expression is 
\begin{eqnarray}
S_{NA}(\mathbf{q}) =  & \frac{\sigma_1}{1 + \xi_{\|}^2(q_{\|} - q_{\|}^0)^2 + \xi_{\bot}^2q_{\bot}^2 + c\xi_{\bot}^4q_{\bot}^4} \nonumber \\
 +  & \frac{\sigma_2}{[1 + \xi_{\|}^2(q_{\|} - q_{\|}^0)^2 + \xi_{\bot}^2q_{\bot}^2 + c\xi_{\bot}^4q_{\bot}^4]^2} 
\label{N_to_A}
\end{eqnarray}

\noindent which is powder averaged and convoluted with the measured resolution
function. In Eq.~(\ref{N_to_A}), $\sigma_1$ and $\sigma_2$ are the amplitudes of the
thermal and random-field terms respectively; $\xi_{\|}$ and $\xi_{\bot}$ are the
correlation lengths parallel and perpendicular to the layer normal; likewise $q_{\|}$
and $q_{\bot}$ are the components of the wave vector $\mathbf{q}$; $q_{\|}^0$ is the
peak position and $c$ is the coefficient of the fourth order correction term. The
amplitude of the random-field term increases as the temperature is reduced
(Fig.~\ref{figure3}).
\begin{figure}
\includegraphics[scale=0.45]{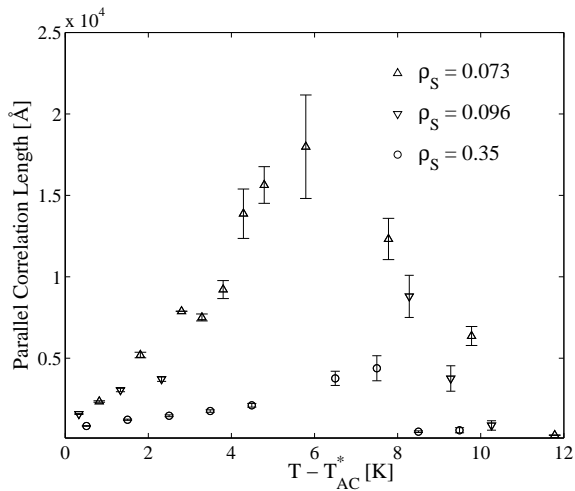}
\caption{\label{figure4} The parallel correlation length, $\xi_{\|}$, for SmA 
order as a function of temperature as a result of fits of Eq.~(\ref{N_to_A}) 
to the data. The effective transition temperature $T_{AC}^*$ is defined
in the text related to Fig.~\ref{figure6}, and the values are listed in
Table~\ref{tab:Tna}. Note that for the pure $\overline{8}$S5 material 
used in this work $T_{NA}^0 - T_{AC}^0 = 7$\,K, and the 
$T_{NA} - T_{AC}^*$ value for all $\overline{8}$S5-aerosil samples is
expected to have almost the same value~\cite{H_and_G}.}
\end{figure}

As with previous studies~\cite{Park,Leheny,Clegg}, it is assumed that the dependence of
$\xi_{\bot}$ and $c$ on $\xi_{\|}$ remains the same in the aerosil samples as it is in
the pure materials. It should be noted that the asymmetric line shape is a consequence
of powder averaging Eq.~(\ref{N_to_A}) for a highly anisotropic correlation volume 
together with a small value for $c$ in
$\overline{8}S5$ which, in turn, means that there are long tails in the transverse 
direction. 
This $\overline{8}S5$ line shape
cannot be reproduced 
using the approximation scheme described in Ref.~\cite{Aerogel_Science} which
does not incorporate explicitly the bulk liquid crystal structure factor. 
Figure~\ref{figure4} shows the
parallel correlation length, $\xi_{\|}$, values extracted from the fits. 
At high temperatures the short-ranged smectic correlations develop in
the nematic phase in a similar manner to that in pure $\overline{8}S5$ and to
other liquid crystal-aerosil gels. At temperatures below the bulk N - SmA
transition temperature the behavior
differs from that of the pure material.
Estimates of the correlation lengths parallel and perpendicular
to the layer normal close to the N - SmA pseudotransition~\cite{res_note} are 
listed for each sample 
in Table~\ref{tab:Tna}.
The parallel correlation lengths are very much greater than those for
8CB-aerosil~\cite{Park, Leheny} and 8OCB-aerosil~\cite{Clegg} at the same $\rho_S$ values.
These correlation lengths are very long compared to the size of the pores, more 
than ten times longer for most samples. The highly anisotropic
correlation volume is evident from the perpendicular correlation length values. These values are
comparatively short, roughly half the mean pore size (Table~\ref{tab:Tna}).
At lower temperatures the behavior in the SmA regime is qualitatively very different from 
that observed for 8CB-aerosil and 8OCB-aerosil systems~\cite{Leheny, Clegg}. The size of the 
correlated SmA regions in a given $\overline{8}S5$-aerosil sample decrease markedly as the 
temperature is lowered toward the SmA - SmC transition whereas $\xi_{\|}$ is roughly
independent of temperature in the SmA region for the materials with no SmC
phase~\cite{Leheny,Clegg}.
Close to the SmA-SmC transition there is a tilt 
instability which can be induced by compression
and this may make the liquid crystal less resistant to the disordered
environment~\cite{Ribotta74}.
\begin{figure}[b]
\includegraphics[scale=0.4]{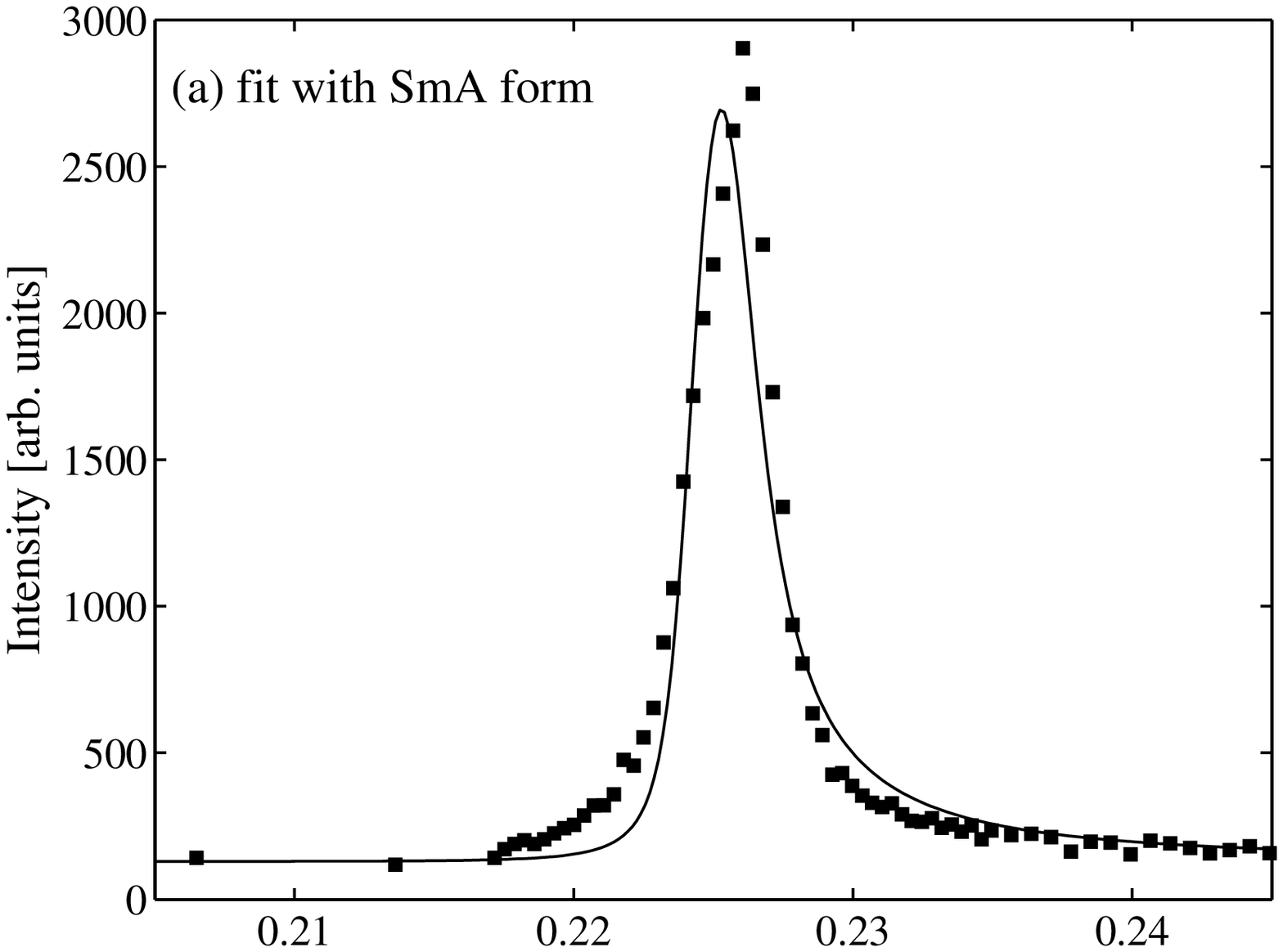}
\includegraphics[scale=0.4]{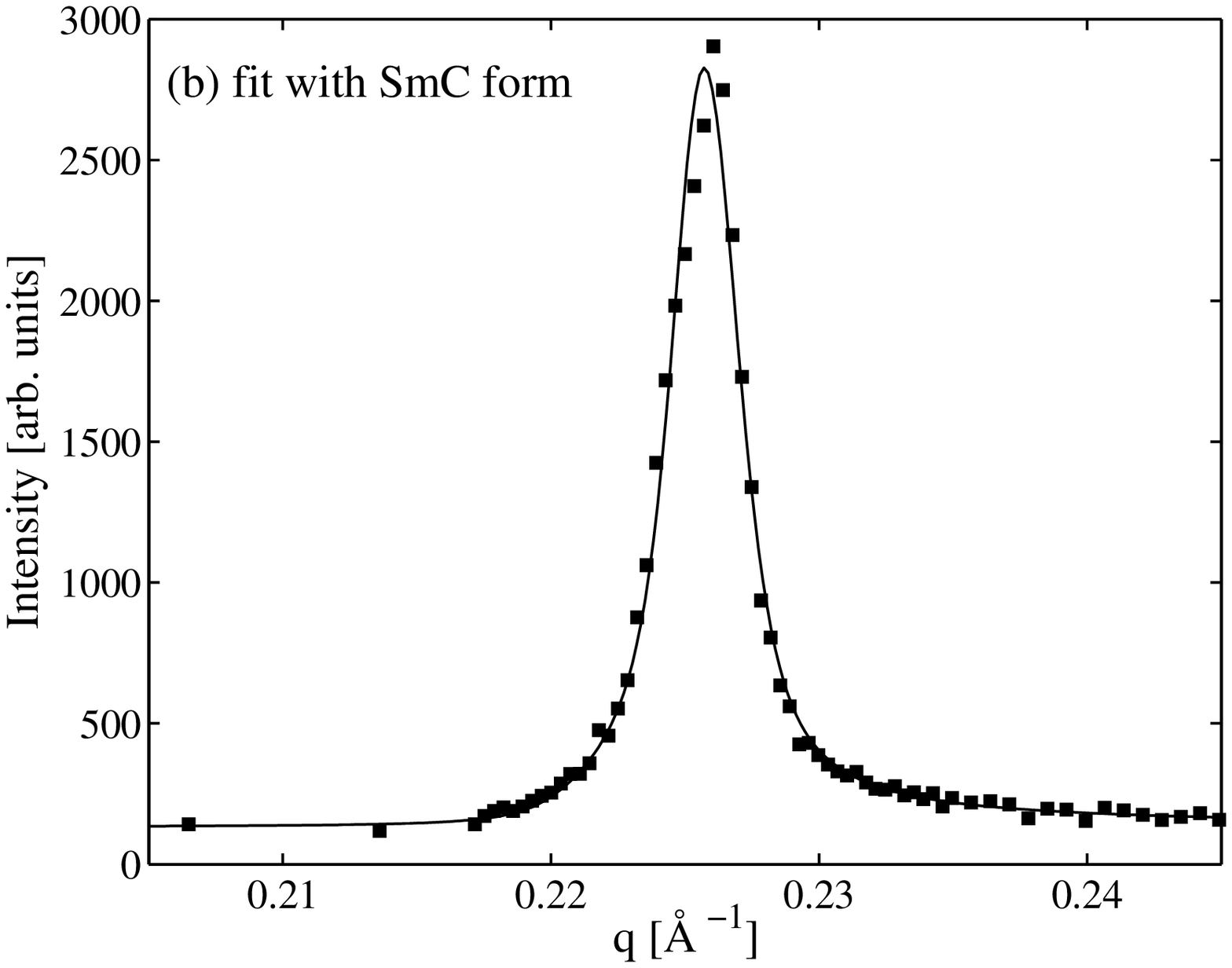}
\caption{\label{figure5} The line shape for a sample with $\rho_S = 0.073$~g~cm$^{-3}$ at
$T = 321.5$ K $< T_{AC}^*$.
The solid line in (a) shows the result of a fit to a powder average of the SmA form given
by Eq.~(\ref{N_to_A}) and that in (b) shows the result of a fit to a powder average of the SmC
line shape given by Eq.~(\ref{A_to_C}).} 
\end{figure}

Figure~\ref{figure5} illustrates the low temperature line shape for the finite range SmC ordered state when
$\rho_S = 0.073$. The asymmetry is pronounced and is different from that observed when the
structure factor has SmA character. The solid line in Fig.~\ref{figure5}(a) is the result of a fit of
the data to the model given in Eq.~(\ref{N_to_A}). This model does not describe the line shape at all
well. The solid line in Fig.~\ref{figure5}(b) is the result of a fit using the SmC
structure factor as the basis for the two component line shape. The SmC structure factor
is taken from studies of the pretransitional N-SmC fluctuations in
$\overline{7}S5$-$\overline{8}S5$ mixtures, the so called Chen-Lubensky structure factor~\cite{MM87}: 
\begin{eqnarray}
S_{AC}(\mathbf{q}) = \frac{\sigma_1}{1 + \xi_{\|}^2(q_{\|} - q_{\|}^0)^2 - U_p q_{\bot}^2 + V_p q_{\bot}^4} \nonumber \\
 + \frac{\sigma_2}{[1 + \xi_{\|}^2(q_{\|} - q_{\|}^0)^2 - U_p q_{\bot}^2 + V_p q_{\bot}^4]^2} 
\label{Chen_Lubensky}
\end{eqnarray}

\noindent with $U_p, V_p > 0$. This is identical to Eq.~(\ref{N_to_A}) except that the 
coefficient of
the $q_{\bot}^2$ term is assumed to be negative. Since the line shape and the
peak position have quite different responses to the porous environment it is more convenient to
restate this structure factor displaying explicitly the component of the peak position 
perpendicular to the long-axis of the molecules, $q^0_{\bot}$:
\begin{eqnarray}
S_{AC}(\mathbf{q}) = \frac{\sigma_1}{1 + \xi_{\|}^2(q_{\|} - q_{\|}^0)^2 + D_p^2(q_{\bot}^2 -
(q_{\bot}^0)^2)^2} \nonumber \\
 + \frac{\sigma_2}{[1 + \xi_{\|}^2(q_{\|} - q_{\|}^0)^2 + D_p^2(q_{\bot}^2 -
(q_{\bot}^0)^2)^2]^2} 
\label{A_to_C}
\end{eqnarray}

\noindent This is seen to give a
good account of the data in Fig.~\ref{figure5} and Fig.~\ref{figure2}(d \& e). The best fit between this
structure factor and the data was found by varying the thermal amplitude $\sigma_1$, the
random-field amplitude $\sigma_2$, the parallel correlation length 
$\xi_{\|}$, the perpendicular
parameter $D_p$, and the peak position $q_c = \surd((q_{\|}^0)^2 + (q_{\bot}^0)^2) 
= 2 \pi / d_c$.
The two components, $q_{\|}^0 = q_c cos{\phi}$ and $q_{\bot}^0 = q_c sin{\phi}$ were 
determined using the previously determined relationship~\cite{Safinya80} between 
the tilt angle, $\phi$, and 
the layer spacing in the SmA phase, $d_a$, that is $\phi = 1.2 \surd(2 (1 - d_c / d_a))$.
\begin{figure}[b]
\includegraphics[scale=0.45]{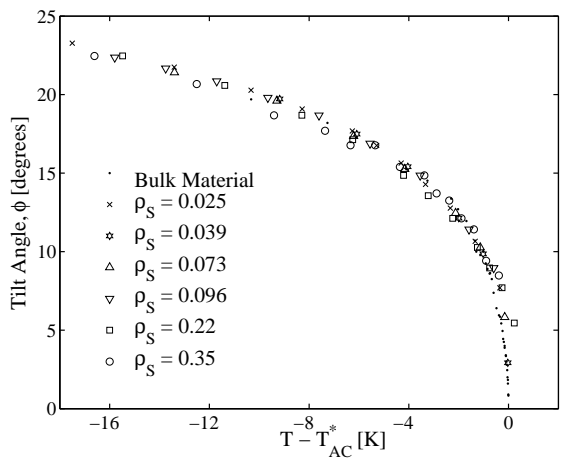}
\caption{\label{figure6} Temperature dependence (and $\rho_S$ independence) of the
molecular tilt angle with respect to the layer normal below $T_{AC}^*$. The data for
bulk $\overline{8}S5$ is taken from Ref.~\cite{Safinya80}. The values of the
effective transition temperatures are given in Table~\ref{tab:Tna}.}
\end{figure}
\begin{figure}
\includegraphics[scale=0.4]{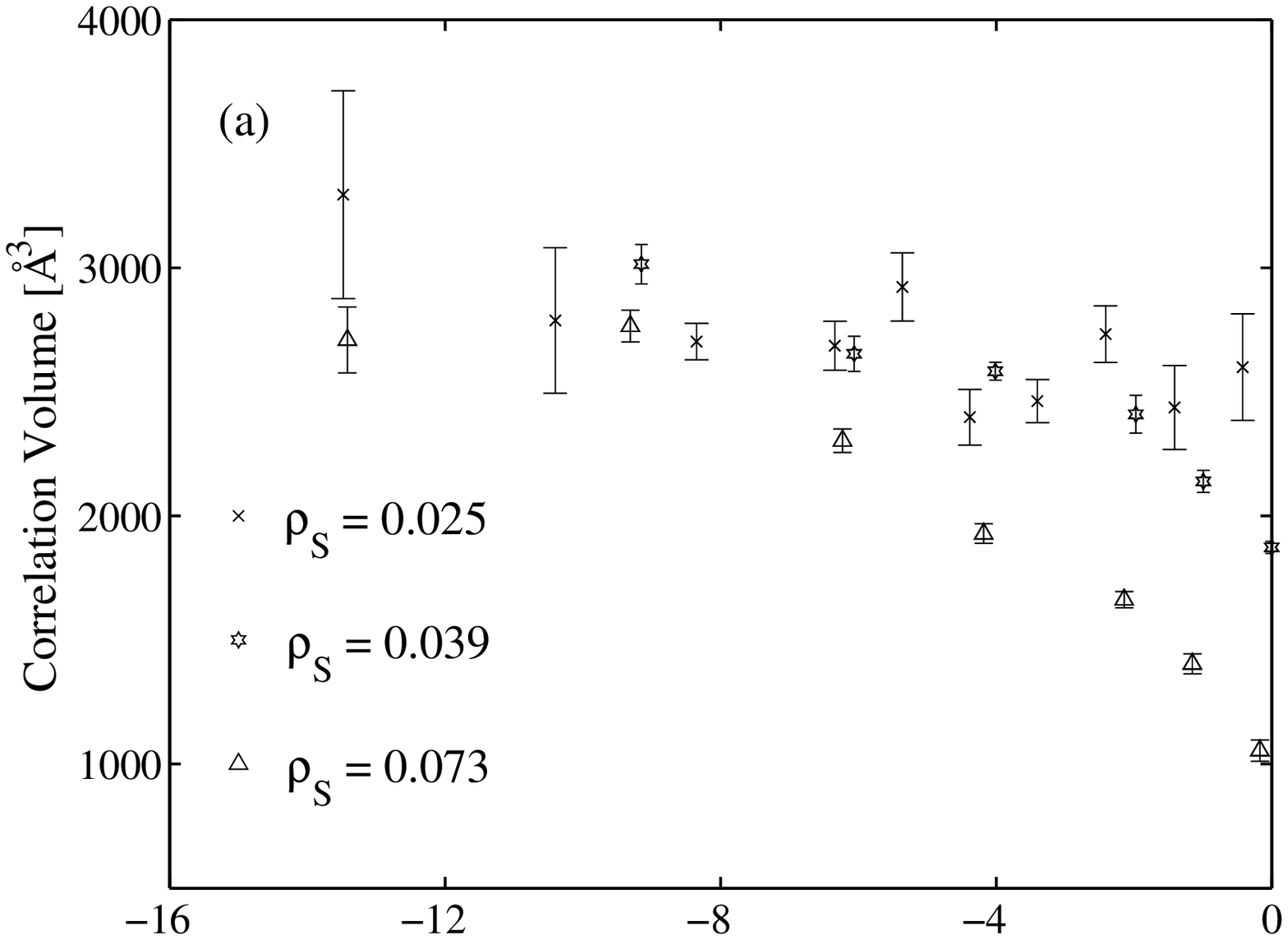}
\includegraphics[scale=0.4]{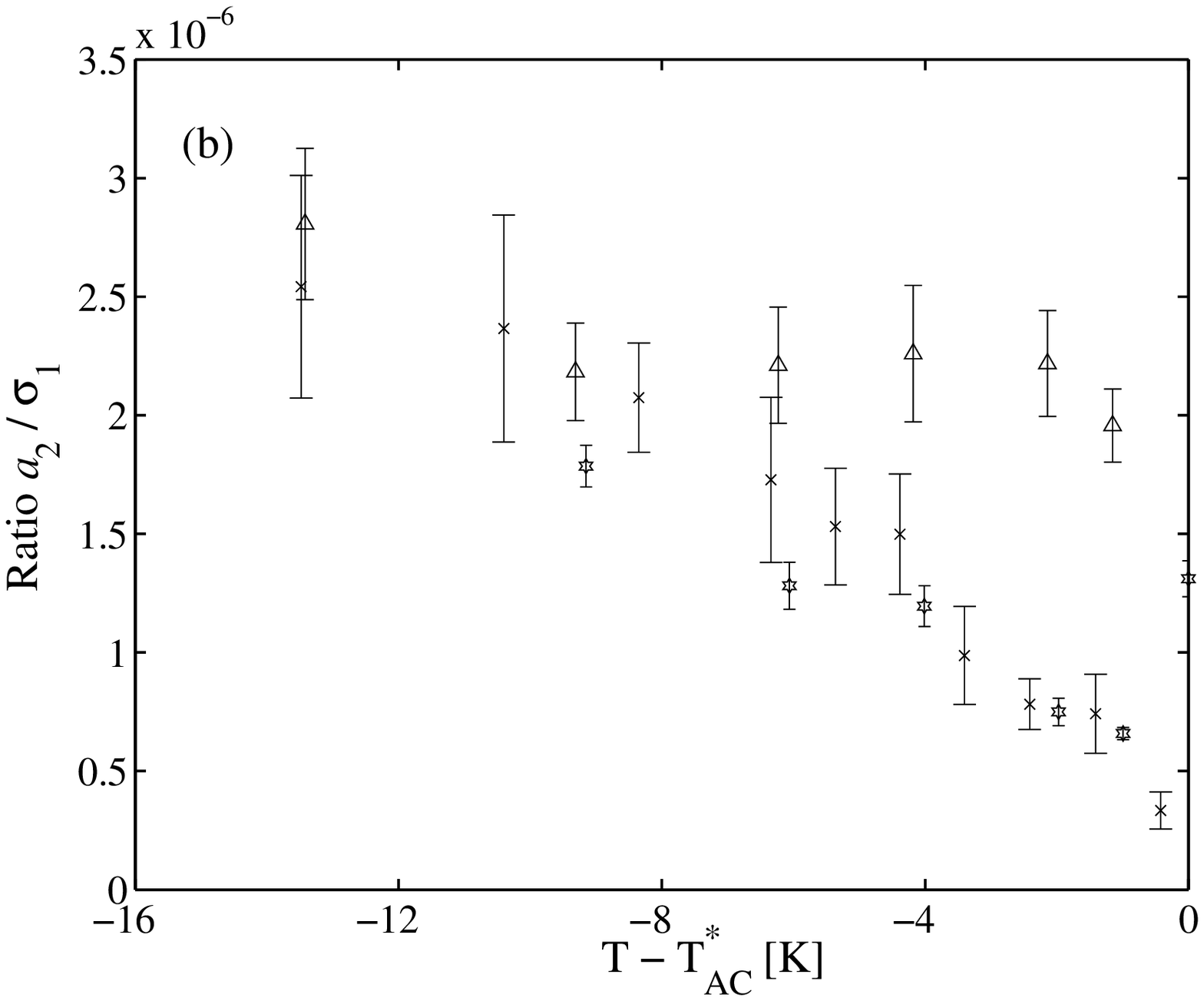}
\caption{\label{figure7} (a) Correlation volume for SmC order as a function of temperature as a 
result of fits of Eq.~(\ref{A_to_C}) to the data. (b) The ratio of the amplitude of the random-field term, 
$a_2$, to the thermal term, $\sigma_1$.}
\end{figure}

The SmA-SmC transition in $\overline{8}S5$ is a mean-field transition with the tilt angle as
the order parameter~\cite{Safinya80}. Hence the pretransitional fluctuations are
weak in the pure material. Figure~\ref{figure6} shows the tilt angle, as a 
function of temperature, for the full range of disorder strength
studied. The tilt angle was determined from the SmC
peak position, $q_c$. The effective transition
temperature, $T_{AC}^*$, was found by fitting the Landau-Ginzburg tricritical
model~\cite{Birgeneau83}
to the temperature dependence of the tilt angle. When plotted versus $T
- T_{AC}^*$, the tilt angles for all samples collapse onto a single
  curve (Fig.~\ref{figure6}).
The robustness of $\phi$ in gels suggests that, while the disorder may be coupling to the
position and presumably the orientation of the smectic layers, it is not strongly affecting the
molecular orientations~\cite{parms_note}. 
Figure~\ref{figure7} shows parameters extracted from the fits of
Eq.~(\ref{A_to_C}) to the data acquired below $T_{AC}^*$. Only results for low $\rho_S$ are presented. In Fig.~\ref{figure7}(a) 
the correlation volume, $\xi_{\|} D_p$, is
shown as a function of temperature in the SmC region. Increasing the aerosil concentration
progressively decreases the extent of the SmC correlations. For higher $\rho_S$ the line shape becomes
distorted as will be described below. In Fig.~\ref{figure7}(b) the ratio of the
random field to thermal amplitudes is presented in the form $a_2 / \sigma_1$, where 
$a_2 = \sigma_2 / \xi_{\|} D_p$. The ratio $a_2 / \sigma_1$ is independent of normalization and 
is predicted to 
have a power law dependence on $\rho_S$~\cite{Germano2}. Increasing aerosil concentration gives 
rise to a
larger random-field contribution to the structure factor. Note that both Figs.~\ref{figure7}(a) and
\ref{figure7}(b) show $\rho_S$ dependence for temperatures just below $T_{AC}^*$ that diminishes at
lower temperatures. The low temperature behavior could be an early symptom of the 
distortion that becomes clear for the higher
density samples.
\begin{figure}
\includegraphics[scale=0.4]{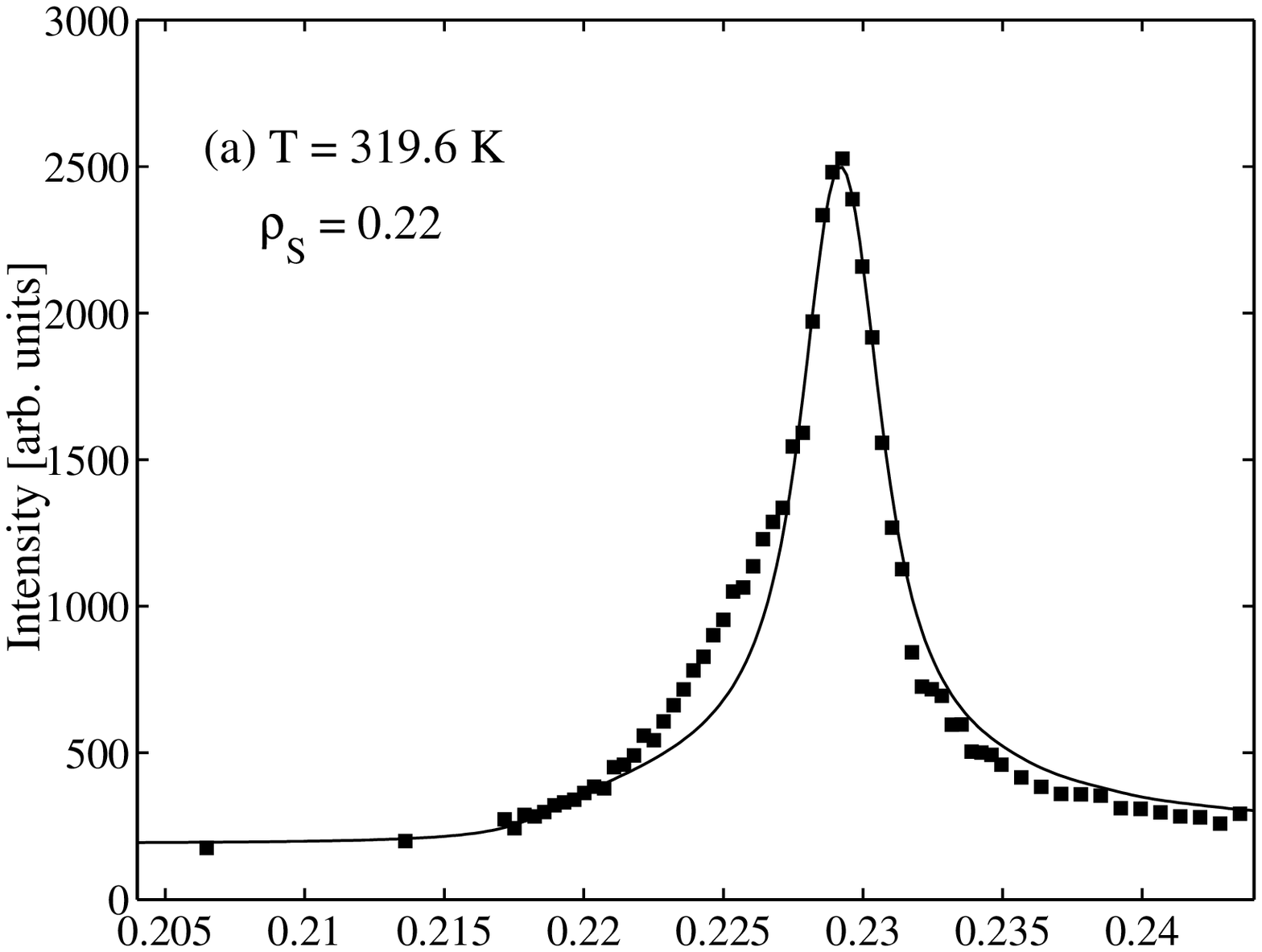}
\includegraphics[scale=0.4]{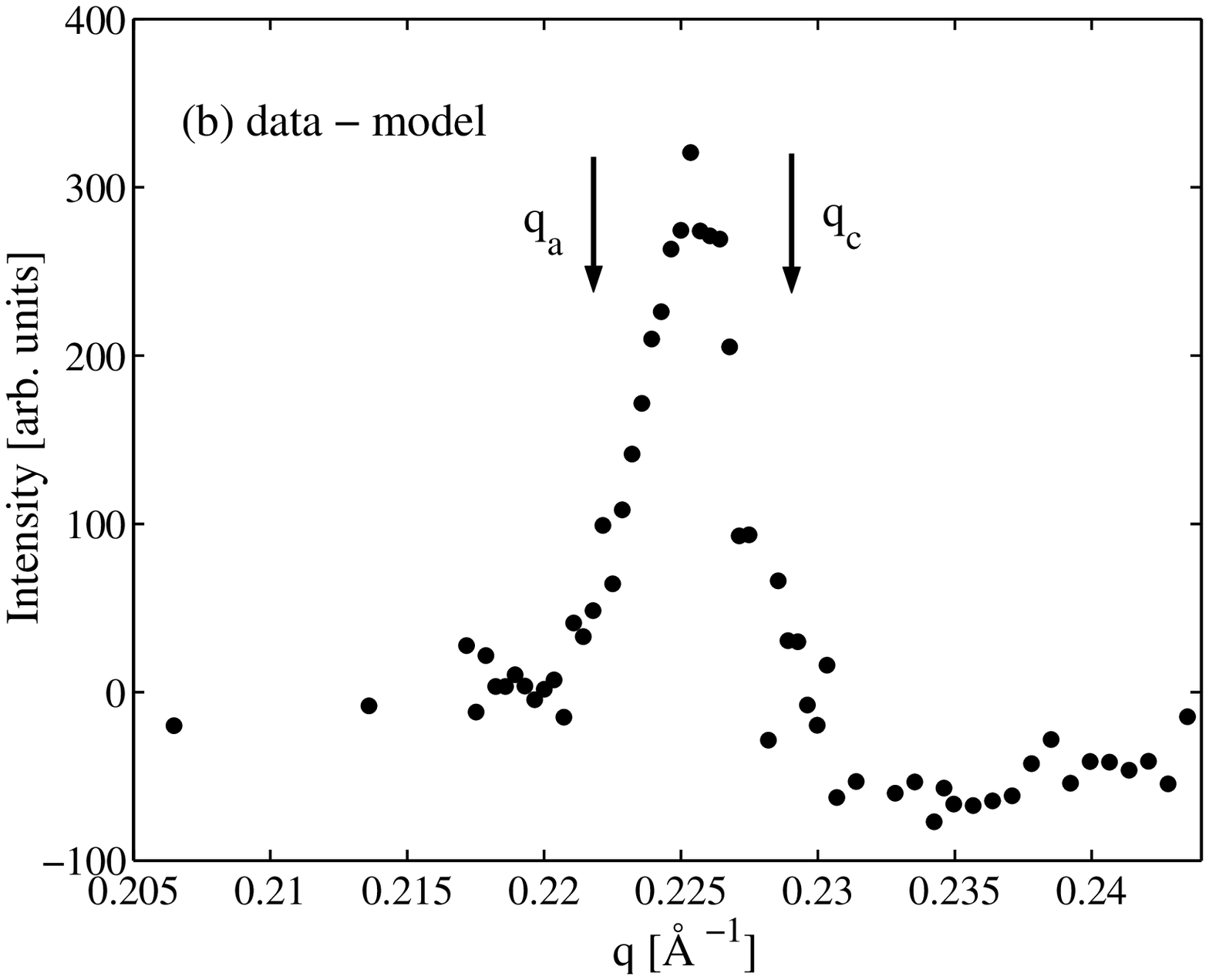}
\caption{\label{figure8} (a) Distorted line shape in the SmC regime at low temperature and high
$\rho_S$. The solid line is the result of a fit of Eq.~(\ref{A_to_C}) to these data. (b) The result of subtracting the fit values from the data points given in part (a). The arrows indicate the positions of the SmA, $q_a$, and SmC, $q_c$, peaks respectively.}
\end{figure}

Table~\ref{tab:Tna} lists the SmA-SmC transition temperatures as a function of $\rho_S$.
There is an initial sharp drop followed by a recovery. This effect is not understood, but
qualitatively similar variations have been observed at N-SmA and N-I transitions in other liquid crystal-aerosil
gels~\cite{Germano1}.

For the highest densities of aerosils and the lowest temperatures the peak shape
corresponding to SmC fluctuations becomes distorted.
A shoulder develops on the low tilt angle side of the reflection peak, as shown in 
Fig.~\ref{figure8}(a). The shoulder becomes more pronounced as
the temperature is reduced and as $\rho_S$ is increased.
Figure~\ref{figure8}(b) shows the result of subtracting the result of a fit (solid
line Fig.~\ref{figure8}(a)) from the data. A peak is observed which falls between the SmA peak
position $q_a = q_{\|}^0$ and the SmC peak position $q_c$. This behavior
indicates that a minority of regions become fixed at lower tilt angles.

\section{Conclusions}
\label{sec:conc}

X-ray diffraction results have been presented for $\overline{8}S5$ aerosil dispersions over
the temperature range for which the bulk liquid crystal has N-SmA and SmA-SmC transitions.
The gel of aerosil particles is observed to destroy the SmA and SmC phases in the sense 
that the ordered states becomes short ranged.
Both the short-range order SmA and SmC structure factors are well described by a
model assuming random fields. The correct thermal fluctuation structure factor has to be used as
the thermal term and its square as the random-field term.
Estimates at low $\rho_S$ and more accurate
values for high $\rho_S$~\cite{res_note} indicate that $\xi_{\|}$, the correlation lengths 
parallel to the layer normal, are very long 
compared to those
of 8CB and 8OCB for the same pore sizes. This suggests that the non-polar molecule 
experiences a weaker pinning field 
than the polar materials do.
The highly anisotropic
correlation volume for this material makes the fitting procedure simpler. It also makes
clear that the two-Lorentzian model used by Bellini and
coworkers~\cite{Aerogel_Science} cannot be applied in this case. That the two component 
line shape gives a good account of data for the SmA - SmC transition in 
$\overline{8}S5$-aerosil gels is intriguing.

For other liquid crystal-aerosil gels exhibiting N - SmA transitions, it has been possible 
to make detailed comparisons
with theoretical models. For
8CB-aerosil~\cite{Park,Leheny,Germano2} and 8OCB-aerosil~\cite{Clegg}, the parameters 
extracted from the line shape compared favorably with the predictions of Ref.
\cite{A_and_P}, giving the
appearance that the gel applies a random field which pins the phase of the smectic density
wave.
The low temperature behavior of the  correlation length, due to enhanced anomalous 
elasticity, 
predicted by Ref.~\cite{R_and_T} has
not been observed for LC-aerosil gels. In $\overline{8}S5$, the SmA regime appears
to be strongly affected by the impending SmA - SmC transition. The correlation
length is never constant, as for a random-fields model, or, increasing on 
cooling as predicted
for anomalous elasticity. Currently there are no theoretical predictions for the
effects of a porous random environment on the SmA - SmC transition.

The SmA-SmC tilt (order parameter) behavior is preserved almost unmodified in the 
aerosil samples. This suggests that any effect that the aerosils may have on the molecular
orientations has little consequence for the transition and that this is true
even for quite high aerosil concentrations. The bare correlation volume for SmA 
materials exhibiting a SmC phase can be estimated from the elastic properties of 
the material and is typically
(70\,\AA)$^3$~\cite{Safinya80}. For the aspect ratio of the $\overline{8}$S5 molecule
this is likely to imply a bare correlation length parallel to the molecule of close
to ten molecular lengths. This long-range of interaction is highly likely to minimize
the influence of random disorder. If some small fraction of the molecules are forced
to point in uncorrelated random directions they will tend to have no net effect:
the large bare correlation length implies an averaging over the disorder.
For high $\rho_S$, the appearance of correlations peaked at a wave vector mid-way 
between those of the SmA and SmC phases indicates that small regions of the 
sample become stuck at a lower tilt angle.

\section{Acknowledgments}

We are grateful to S.~LaMarra for technical assistance, to N.A.~Clark
for discussing unpublished research, and to M.~Ramazanoglu for helpful comments. Funding in Toronto was
provided by the Natural Science and Engineering Research Council and
support was provided at WPI by the NSF under award
DMR-0092786 and at JHU under award DMR-0134377. Dr Mary E. Neubert and David G. Abdallah Jr. 
were supported by
NSF grant DMR 89-20147.


\begin{thebibliography}{100}
\bibitem{res_note}
All correlation length values listed in Table~\ref{tab:Tna} were determined by
interpolating at a temperature $T = T_{AC}^* + 7$\,K (based on the assumption,
supported by Ref.~\cite{H_and_G}, that the SmA range has the same value in all
aerosil samples as in pure $\overline{8}$S5), and they thus
reflect the 
likely behavior close to the N - SmA pseudotransition. The values for the $\rho_S
= 0.025, 0.039, 0.073$ samples are based on fits to the tails of the peak, due to
the resolution limited width.
\bibitem{parms_note}
The results of fits of the tricritical to mean-field crossover model to the 
temperature dependence of the tilt
angle yields variations in the fit parameters as $\rho_S$ is increased. The fourth
order term in the free energy expansion appears to increase with respect to the 
second order term with increasing aerosil concentration. This is consistent with the
system moving away from tricriticality. The variation is small and there is no
obvious trend in the raw data.
\end{thebibliography}
\end{document}